\documentclass[10pt,letterpaper]{article}
\usepackage[top=0.85in,left=2.75in,footskip=0.75in]{geometry}

\usepackage{changepage}

\usepackage[utf8x]{inputenc}

\usepackage{textcomp,marvosym}

\usepackage{fixltx2e}

\usepackage{amsmath,amssymb}

\usepackage{cite}

\usepackage{nameref,hyperref}
\usepackage[right]{lineno}

\usepackage{microtype}
\DisableLigatures[f]{encoding = *, family = * }

\raggedright
\usepackage[aboveskip=1pt,labelfont=bf,labelsep=period,justification=raggedright,singlelinecheck=off]{caption}

\makeatletter
\renewcommand{\@biblabel}[1]{\quad#1.}
\makeatother

\date{}

\usepackage{lastpage,fancyhdr,graphicx}
\usepackage{epstopdf}
\fancyheadoffset[L]{2.25in}
\fancyfootoffset[L]{2.25in}

\begin{document}
\vspace*{0.2in}

\begin{flushleft}
{\Large
\textbf\newline{Positive Feedback and Synchronized Bursts in Neuronal Cultures} 
}
\newline
\\
Yu-Ting Huang\textsuperscript{1,2},
Yu-Lin Chang\textsuperscript{2},
Chun-Chung Chen\textsuperscript{2},
Pik-Yin Lai\textsuperscript{1},
C. K. Chan\textsuperscript{1,2}
\\
\bigskip
\textbf{1} Dept. of Physics and Center for Complex Systems, National Central University, Chungli, Taiwan 320, ROC
\\
\textbf{2} Institute of Physics, Academia Sinica, Nankang, Taipei, Taiwan 115, ROC
\\
\bigskip

%
%


\textcurrency Current Address: Institute of Physics, Academia Sinica, Taipei, Taiwan 



* pylai@ncu.phy.edu.tw and ckchan@gate.sinica.edu.tw

\end{flushleft}
\section*{Abstract}
Synchronized bursts (SBs) with complex structures are common in neuronal cultures. Although the origin of SBs is still unclear, they have been studied for their information processing capabilities. Here, we investigate the properties of these SBs in a culture on multi-electrode array system. We find that structures of these SBs are related to the different developmental stages of the cultures. A  model based on short term synaptic plasticity, recurrent connections and astrocytic recycling of neurotransmitters has been developed successfully to understand these structures. A phase diagram obtained from this model shows that networks exhibiting SBs are in an oscillatory state due to large enough positive feedback provided by synaptic facilitation and recurrent connections. In this model, the structures of the SBs are the results of intrinsic synaptic interactions; not information stored in the network.



\section*{Introduction}

Synchronized bursts (SBs) \cite{MEABurst} are common in our brains. They originate from the collective dynamics of neurons in the neural networks. These bursts can be related to the normal functioning of the brain or to some pathological states such as epilepsy. For the normal functions, it has been proposed that these collective dynamics might encode information and perform computation. In fact, neuronal cultures grown on top of a multi-electrode array (MEA), which invariably generate SBs, have become a standard experimental platform for the study of computational capabilities of living neural networks. For example, Ben-Jacob \textit{et al.} studied these SBs for encoded information \cite{HiddenCor}, complexities and even memories capabilities \cite{complexity} while Lee \textit{et al.} attempted to train \cite{Lee2012} these SBs by using external stimulations. The goal of these kind of works is to understand the mechanisms of these SBs and ultimately to construct a living neural chip \cite{chip} with useful functions. Unfortunately, very little success has been achieved towards this goal despite of the extensive efforts in the last two decades. 

One of the main difficulties for the modeling of SBs is that the dynamics of the SBs can vary widely depending on the conditions of the cultures \cite{MEABurst}. This rich repository of dynamics might be the results of memories, learning, etc., in the network as mentioned above. However, it might also simply be the complex behaviors generated by simple dynamical systems \cite{pattern}. It is known that SBs are controlled by synaptic transmission \cite{cohen2008} and occur only when there are enough connections \cite{CKC2004} in the network.   Intuitively, minimal recurrent connections are needed for the re-excitation (positive feedback) of the network to maintain a burst while synaptic mechanism should determine the detailed dynamics of a SB.  Since even simple dynamical systems with different amount of positive feedback can lead to complex behaviors, it is plausible that SBs observed are simply the results of different amount of re-excitations.

In this article, we report our investigation by experiments and modeling of a neuronal network grown on a MEA to understand the origin of these SBs. Specifically, we are able to produce a special form of SBs similar to reverberations \cite{Bi2005} or superbusrts \cite{MEABurst} by controlling the neuron densities in the cultures. The dynamics of the SBs are characterized by the firing-rate-time-histogram (FRTH, see Fig. \ref{raster} ); with the shapes of these FRTH corresponding consistently well to different stages of development of the network. A mean field model based on a recurrent connection $J$ and short term synaptic plasticity (STSP) is constructed to reproduce these measured FRTHs. We find that the STSP mechanism \cite{mongillo2008} is able to reproduce generic features of the FRTHs only when an additional recycling of neurotransmitters mechanism \cite{Hennig2013} characterized by the baseline level of the available neurotransmitter ($X_0$) is included. A phase diagram in terms of $J$ and $X_0$ shows that networks exhibiting SBs are in an oscillatory state due to the positive feedback provided by synaptic facilitation and recurrent network structure. Our finding suggests that, at least for the types of SBs studied in our experiments, the occurrence of SBs just signals that there are too many connections in the network; with no relation to information stored in the network.

\section*{Materials and Methods}
\subsection*{Cell culture.}

Neuronal cultures grown on top of multi-electrode arrays (MEA) are used in our experiments. For cultures preparation\cite{CKC2004}, cortex tissues are extracted from embryonic day 18 (E18) Wistar rat embryos. The tissues are digested by 0.125\% trypsin and gentle triturated by a fire-polished Pasteur pipette to isolate cells. A small drop (5 $ \mu $L) of cell suspension is added on the MEA (MEA60-200-ITO, Qwane Biosciences) that has been pre-treated with 0.1\% Polyethylenimine, yielding a density of 4 $ \times $ $ 10^{4} $ cells/ $\mathrm{mm}^2$. The MEAs are filled with culture medium (DMEM with 5\% FBS, 5\% HS and 1\% penicillin/streptomycin) 30 min after seeding.  Samples are incubated at 37 \textdegree C with 5\% CO\textsubscript{2} and half of the medium is changed twice a week.

\subsection*{Firing activities recordings.}
Firing activities from the cultures are recorded extracellularly by 60 electrodes (ITO transparent electrodes with 40 $\mu$m diameter and 200 $\mu$m spacing, arrangement of $8 \times 8$ grids without 4 corners) in a MEA 1060-Inv-BC (Multi Channel systems) with 1100X amplification at a sampling rate of 20 kHz. Firing activities of the cultures are recorded at 33 \textdegree C with 5\% CO\textsubscript{2} by using MC\_RACK software (Multi Channel Systems). Before recording, samples are kept in the recording condition for 10 minutes for adaptation. Data are recorded with samples in culture medium (CM) \cite{CKC2004} for 10 minutes; except for experiments of magnesium free samples which are kept in balanced salt solution (BSS containing (in mM): 130 NaCl, 5.4 KCl, 1.8 CaCl\textsubscript{2}, 5.5 Glucose, 20 HEPES with different concentration of MgCl\textsubscript{2}) \cite{CKC2004}. 

\subsection*{Burst and subburst detection.}
The firing activities recorded by the MEA are analyzed by a custom made software written in MATLAB (The Math-Works, Natick, MA). Low frequency components of the data are filtered by high-pass filtering at 200 Hz to remove drifts of the signals. Spikes are then detected from the filtered data by using a threshold of five standard deviations of the noise levels for each electrode. A SB is detected if more than 10 spikes occurred within 5 ms and persists for more than 100 ms from all electrodes. 

\begin{figure}[!h]
	\begin{center}
		\includegraphics*[width=\columnwidth]{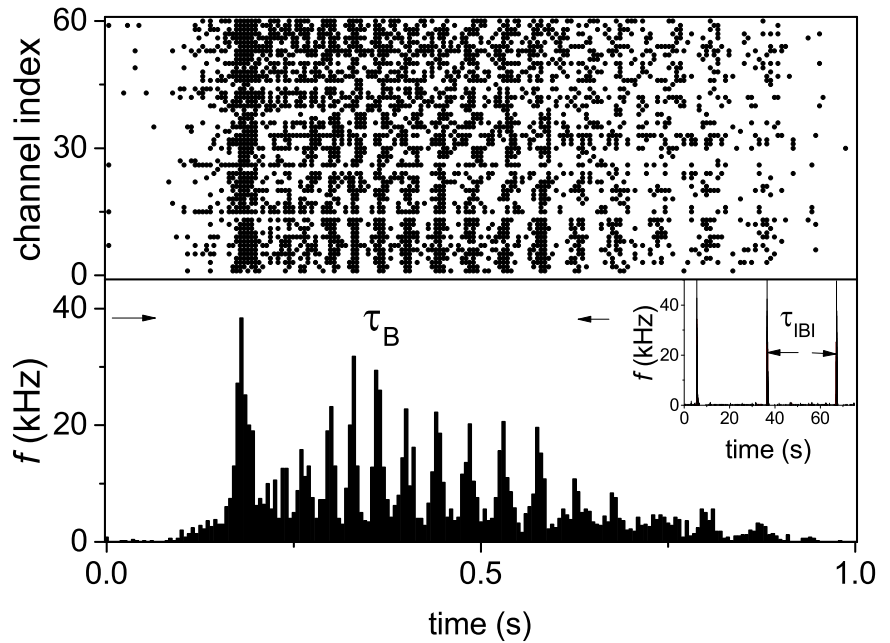}
	\end{center}
\caption{{\bf Raster plot and its firing-rate-time-histogram of a synchronized burst.}
		a) Raster plots of spiking events (a dot) in the 60 channels of the MEA as a function of time.
		b) Firing-rate-time-histogram (FRTH) in a 5 ms time window constructed from a) together with the definition of burst duration ($ \tau_{B} $). The inset shows the FRTH containing 3 SBs together with the	definition of inter-burst interval ($ \tau_{IBI} $).
		}
	\label{raster}
\end{figure}

Fig. \ref{raster} shows a raster plot and a FRTH of a SB from a typical experiment recorded by the MEA, which has a burst duration ($\tau_B$) of the order of one second. From the raster plot, it can be seen that there seems to be sub-bursts within the SB. In order to show these sub-bursts clearly, the FRTH with non-overlapping 5 ms time bin from all the 60 channels is presented in Fig. \ref{raster}b.  Note that, for typical experiments, more than 80\% of the spikes recorded from the cultures are distributed inside the synchronized bursts as shown in the inset of Fig. \ref{raster}b . Therefore,  most the information of the state of the culture is coded into these sub-bursts in a SB.

\subsection*{TMX Model.}
During a SB, most of neurons in the cultures are firing synchronously; suggesting that SB can be treated as a mean-field phenomenon; namely the dynamics of the whole network is similar to the dynamics of a single cell \cite{Zhan2011}. With this picture, the mechanism of these SBs can be understood through the modeling of the mean firing rate $E(t)$ of a single cell. Below, we extend the Tsodyks-Markram (TM) \cite{mongillo2008} model of STSP for this purpose. In the TM model, $E(t)$ of a recurrent network receiving a global inhibition $I_0$ is governed by (Supplementary Materials of \cite{mongillo2008}) :
\begin{equation}
\frac{dE}{dt}=\frac{1}{\tau}\left[-E+\alpha\ln(1+e^{\frac{JuxE+ I_0}{\alpha}})\right]\label{Et}
\end{equation}
\noindent
where $\alpha$ is the threshold of the gain function and $u$ is the release probability of the available neurotransmitter fraction ($x$). Note that the positive feedback ($JuxE$) contains both the structural ($J$) and synaptic factors ($ux$). In the TM model, the dynamics of the depression and facilitation in the synaptic factors are implemented as: 
\begin{equation}
\frac{dx}{dt}=\frac{{\chi_0}-x}{\tau_{D}}-uxE\label{xt}
\end{equation}
\begin{equation}
\frac{du}{dt}=\frac{U-u}{\tau_{F}}+U(1-u)E.\label{ut}
\end{equation}
\noindent
where ${\chi_0}$ and $U$ are the baseline level of $x$ and $u$ respectively. The time scales are: $\tau\sim10$, $\tau_{D}\sim100$ and $\tau_{F}\sim 1000$ ms. In the original TM model, ${\chi_0} = 1 $, with a big enough positive feedback ($J$ and $U$), the TM model can indeed produce oscillations \cite{pnas2013} with time scales similar to that of the sub-bursts shown in Fig. \ref{DIV}. However, these oscillations in the TM model will not stop as in a SB.

In order to reproduce our experimental observations, we need to modulate the amount of the positive feedback to stop the sub-bursts. One could control either $x$ or $u$ through modulating their base values ${\chi_0}$ or $U$ respectively. Since $U$ controls the sub-burst oscillations and the recycling of $x$ will be affected due to repeated firings \cite{Hennig2013}, we choose to control $x$ by introducing a time dependent ${\chi_0}$ which is modeled as:
\begin{equation}
\frac{d{\chi_0}}{dt}=\frac{X_0- {\chi_0}}{\tau_{X}}-\beta E\label{Xot}
\end{equation}
\noindent
for some baseline constant $X_0$, time constant $\tau_X (>>\tau_D)$ and a fatigue rate constant $\beta$. Here one expects the fatigue rate is an even slower process with $\beta<<1/\tau_X$, representing that prolong firing would render significant portion of the neurotransmitters not be available \cite{Zhan2011}. We will refer to this extension as the TMX model. Note that ${\chi_0}$ is similar to the "super-inactive state" used by Volman \cite{Volman} to model reverberations in cultures.

\section*{Results}

\subsection*{Properties of synchronized bursts.}
\begin{figure}[!h]
	\begin{center}
		\includegraphics*[width=\columnwidth]{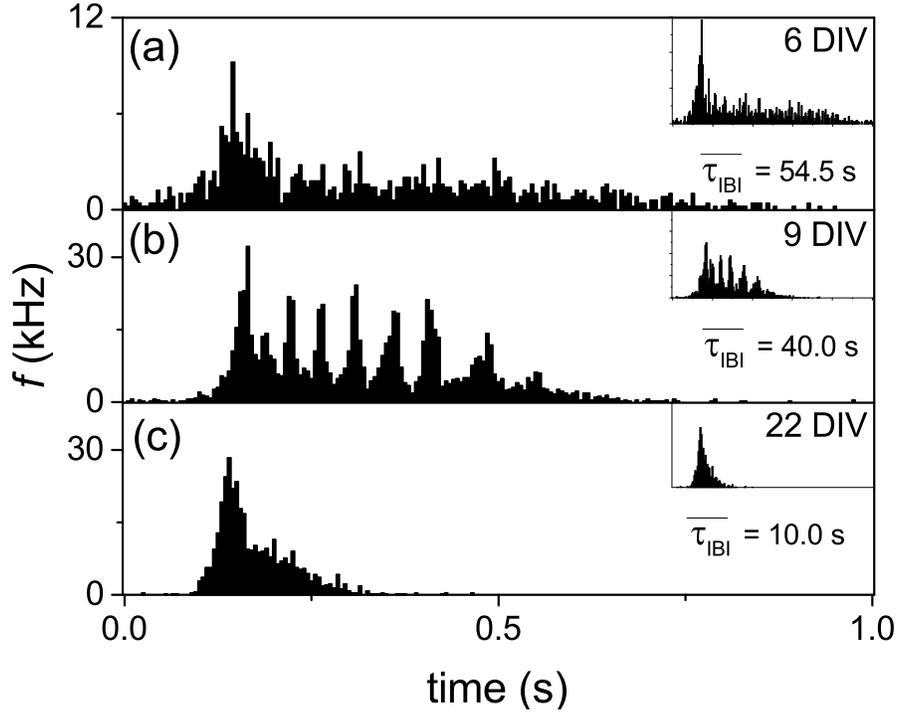}
	\end{center}	
	\caption{ {\bf Structures of FRTH at (a) 6 DIV, (b) 9 DIV and (c) 22 DIV}. The mean $ \tau_{_{IBI}} $($\overline{\tau_{_{IBI}}}$) are also shown in the figures. The insets are the FRTHs of another SB in the same recording. This classification into three types can be observed in 28 out of 30 samples from 14 dissections.}
	\label{DIV}
\end{figure}

Similar to other experiments \cite{MEABurst}, SBs can be observed from the cultures around 7 days after seeding (7 DIV) and characterized by FRTHs.  A remarkable feature of the FRTHs observed is that FRTHs from different SBs measured within 30 minutes from the same sample all have similar features. Fig. \ref{DIV} shows two measured FRTHs from the same sample within 10 minutes as a function of DIV with one of them being shown in the insets. That is: the network mechanism responsible for the SB is just repeating itself during different SBs. Thus, a FRTH constructed from anyone of the SBs can be used to represent the state of the network. 

\begin{figure}[h]
		\begin{center}
			\includegraphics*[width=\columnwidth]{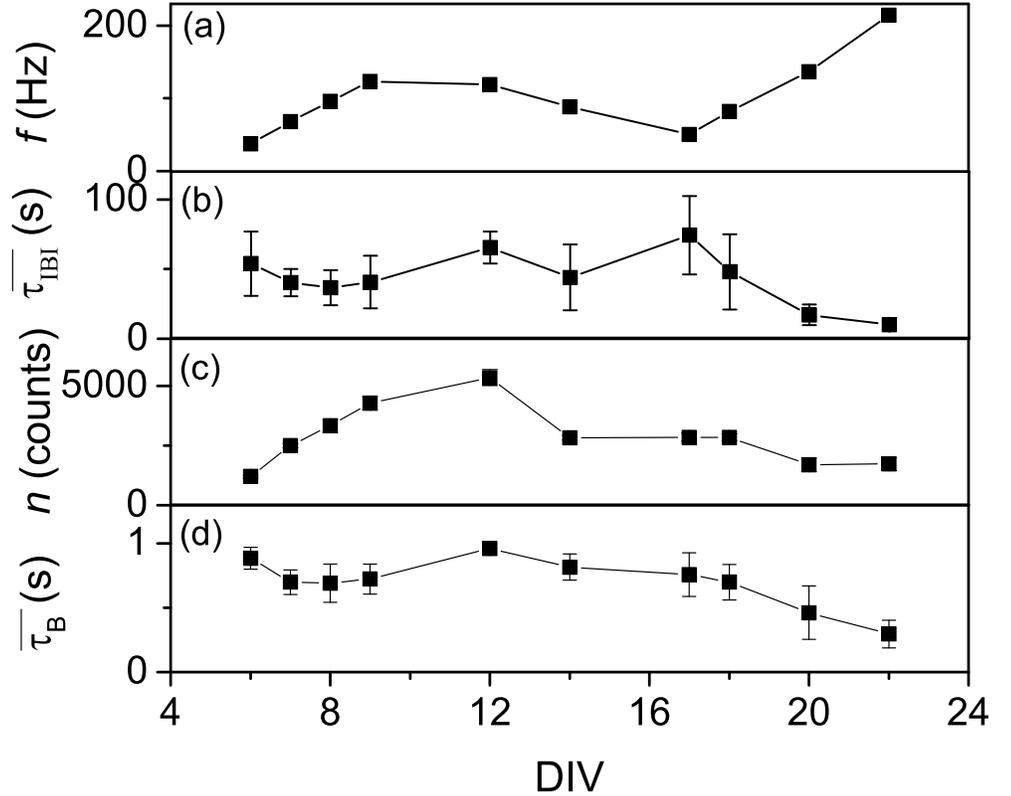}
		\end{center}
	\caption{{\bf Characteristics of SB at different DIVs. }
		a) Mean firing rate ($ f $) within 10 min. The $ f $ at 22 DIV is twice that at 9 DIV.  f is defined as the total number of spikes (N) divided by the recording time. The error (1/$ \sqrt{N} $) is smaller than the symbol. 
		b) Mean inter-burst-interval ($\overline{\tau_{_{IBI}}}$). Note that the $\overline{\tau_{_{IBI}}}$ at 9 DIV is 4 times that at 22 DIV.
		c) Average number of spikes  within SBs ($n$) . Error bars of data are smaller than the symbol.
		d) Mean burst duration $ \overline{ \tau_{B} } $.
		The error bars in the figures are the standard deviation of the data.
	}
	\label{MEAcount}
\end{figure}

The measured FRTHs can be classified into three types based on the shapes of the FRTHs as shown in Fig. \ref{DIV}. In the first type ( $<$ 7 DIV), SBs can be observed with a low firing rate and there is only one single peak in the FRTH. In the second type (7$ <$ DIV $ <$ 20), sub-peaks can be observed within the FRTH. For example, on 9 DIV, seven sub-bursts can be seen in Fig. \ref{DIV}b. At later DIVs, in the third type (DIV$ > $ 20), the sub-peaks disappear and the SB is then consisted only of a single peak similar to that of the first stage but at a higher firing and repetition rates.  The occurrence of the SBs increases from $\sim$0.01 Hz (at 7 DIV) to $\sim$0.1 Hz (at 20 DIV) as the cultures mature but the burst duration and the number of spikes within a SB decrease as DIV increases as shown in Fig. \ref{MEAcount}. 

\subsection*{FRTH under different [Mg$ ^{2+} $] .}
\begin{figure}[h]
	\begin{center}
		\includegraphics*[width=\columnwidth]{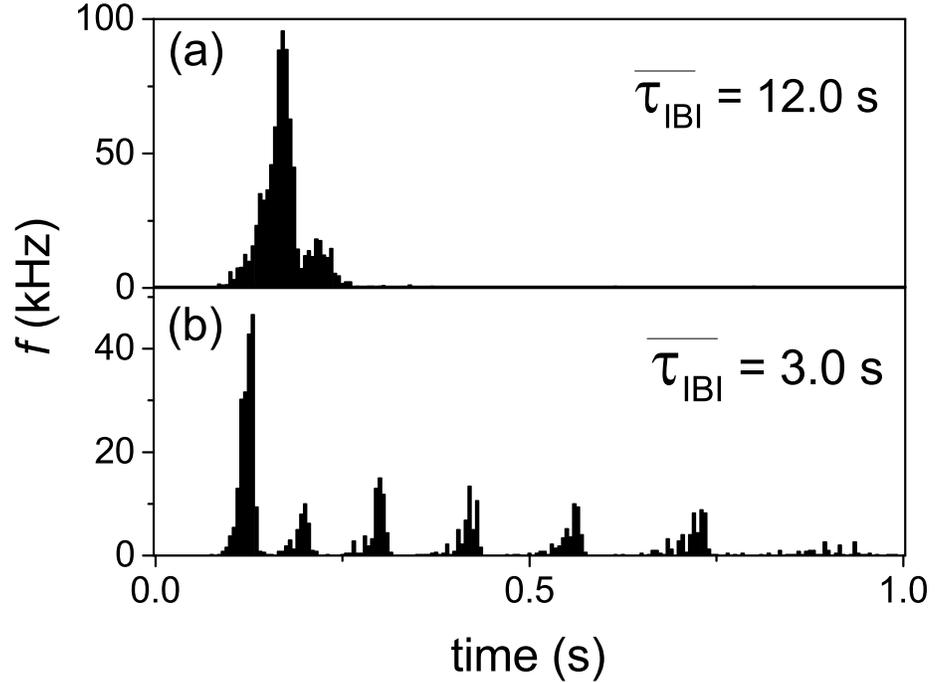}
	\end{center}
	\caption{ {\bf Effect of  [Mg$ ^{2+} $]  on the FRTH at 34 DIV.} (a) In culture medium with 0.8 mM  [Mg$ ^{2+} $] . (b) In BSS with 0 mM  [Mg$ ^{2+} $] . 
		 The induction of sub-burst shown here can be observed in 8 out of 10 samples from 6 dissections.}
	\label{treat}
\end{figure}

Fig. \ref{DIV} shows that different developmental stages of the culture can be represented by FRTHs with different characteristics. To test the effects of synaptic mechanism on the features of measured FRTHs, we have also performed experiments with reduced extracellular magnesium concentrations ( [Mg$ ^{2+} $] ) which can modify the efficacy of synaptic connections through the blocking of the NMDA receptors. Fig. \ref{treat} shows the effects of  [Mg$ ^{2+} $]  on the structure of FRTHs. It shows that sub-bursts within the SB can be induced by decreasing the  [Mg$ ^{2+} $]  from the normal value of 0.8 to 0 mM. The FRTHs shown in Fig. \ref{treat} have similar properties of those shown in Fig. \ref{DIV}; namely FRTHs constructed from different SBs share similar features. Fig. \ref{treat} shows that the features in the measured FRTHs are controlled jointly by the network structure (DIV) and synaptic mechanism.

\subsection*{Properties of TMX model.}
\begin{figure}[h]
	\begin{center}
		\includegraphics*[width=\columnwidth]{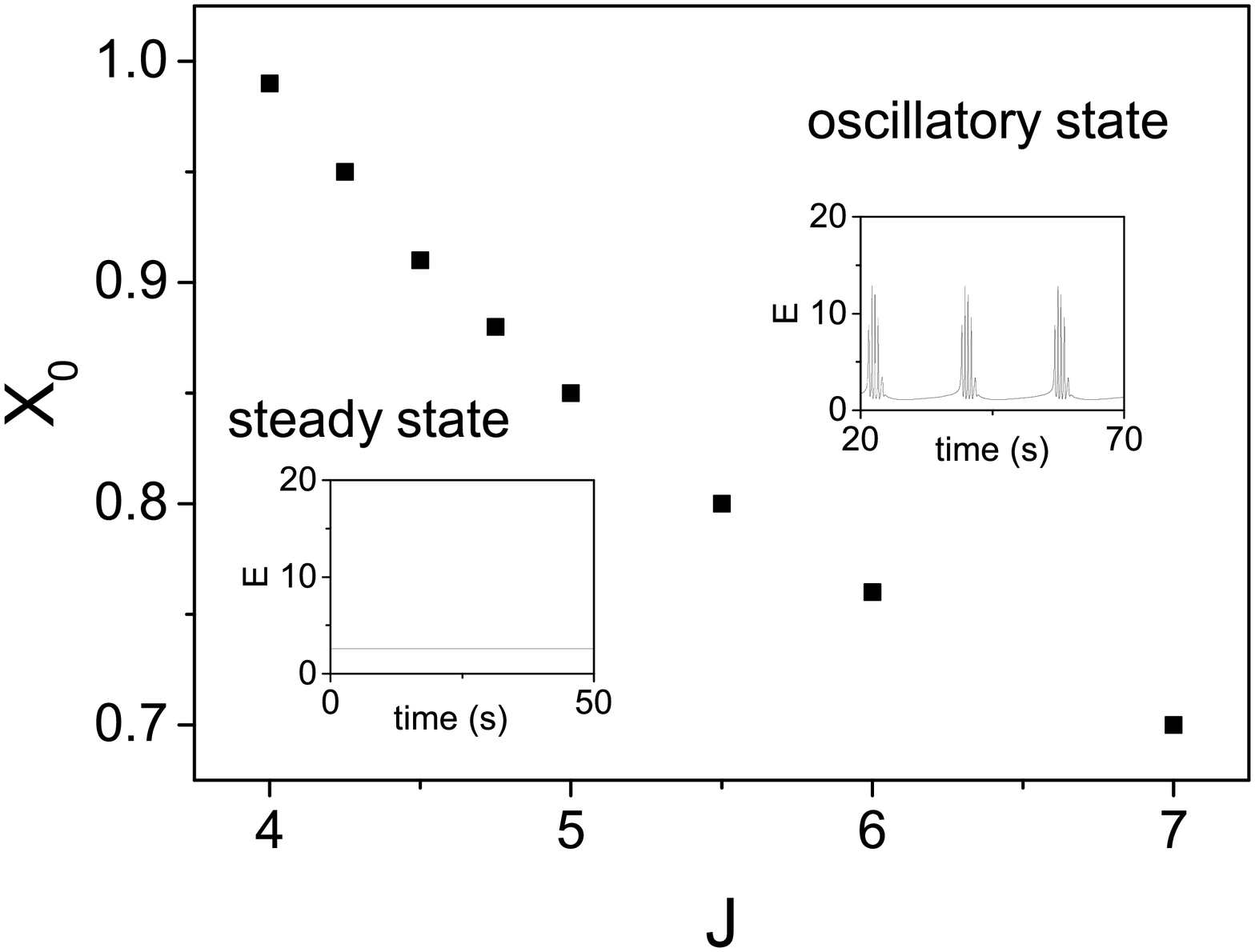}
	\end{center}
	\caption{{\bf Minimal $X_0$ for eliciting an oscillatory firing for a given value of $J$ for $U=0.3$ and $\tau_D=0.15 $ s .} Other parameters are $\tau_X=20$ s, $I_0=-1.3$, $\beta=0.01$, $\tau=0.013$ s, $\tau_F=1.5$ s and $\alpha=1.5$. The insets show the characteristic of $ E(t) $ (in Hz) in the steady and the oscillatory states.  }
	\label{XXvsJ}
\end{figure}

The TMX model has at least two interesting states. The first one is a low firing rate ($E \sim O(1)$) steady state and the other is a periodic state with $E(t)$ showing sub-burst oscillations. This latter state is just the persistent oscillatory state of the TM model now modulated by a time dependent ${\chi_0}$. One can consider the low firing rate state as the cultures in early DIV with little activities and the periodic $E(t)$ as the system at later DIV with SBs. Fig. \ref{XXvsJ} shows a phase diagram for these states in terms of $J$ and $X_0$. Note that the system will be in the periodic state only when either $X_0$ or $J$ (positive feedback) is large enough.

\subsection*{Simulation results to mimic DIV.}
\begin{figure}[h]
	\begin{center}
		\includegraphics*[width=\columnwidth]{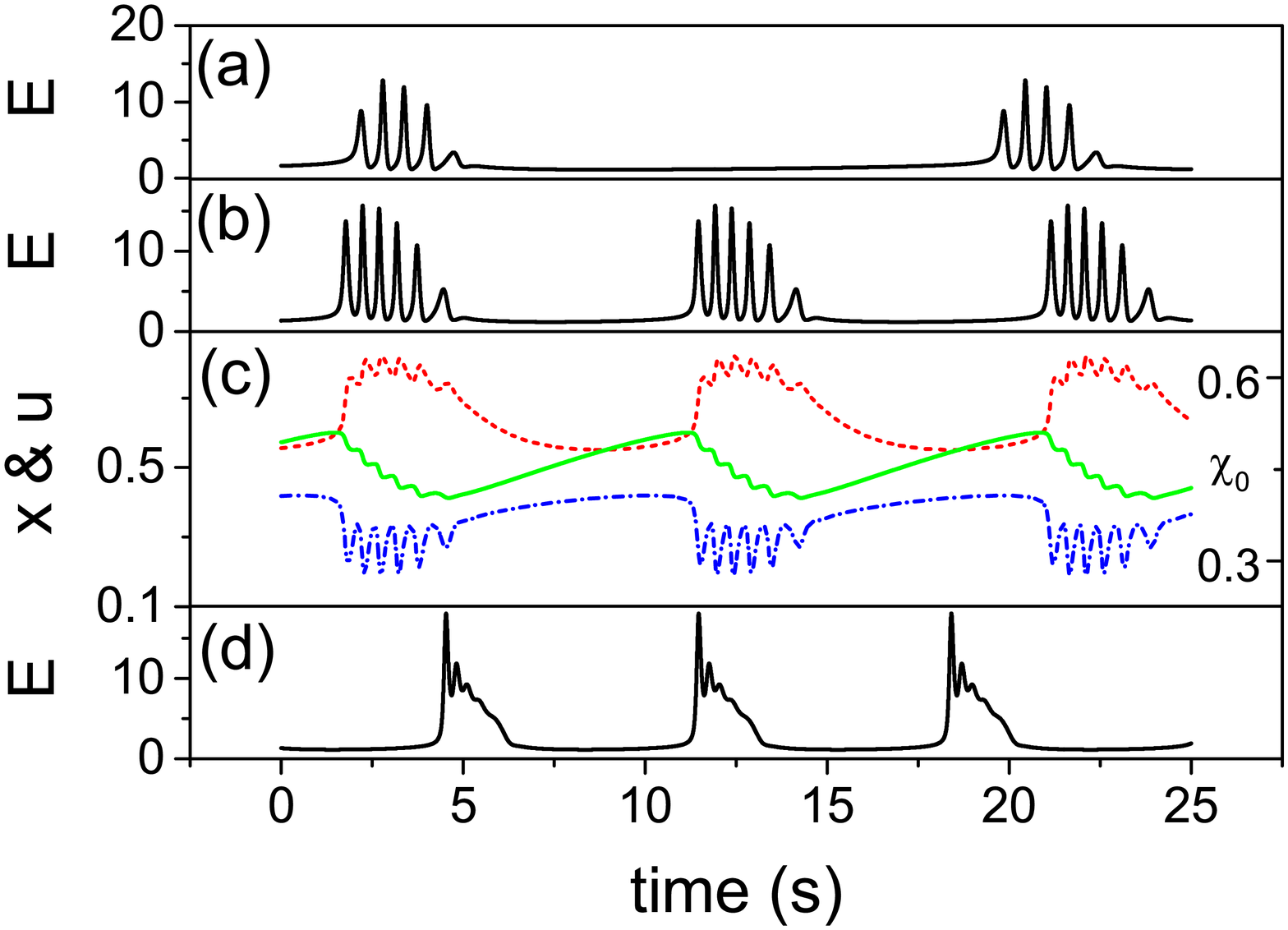}
	\end{center}
	\caption{{\bf Time courses of $E(t)$ (in Hz) of the TMX model to mimic the FRTH at different DIV.}  (a) $J=4.8$, $\tau_D=0.2$ s and $U=0.28$. (b) $J=5.8$, $\tau_D=0.15$ s and $U=0.3$. (d) $J=6.8$, $\tau_D=0.1$ s and $U=0.32$. (c) Corresponding time courses of $ x $ (blue dash-dot), $ u $ (red dot) and ${\chi_0}$ (green) for time course (b). Other parameters are the same as Fig. \ref{XXvsJ}.}
	\label{Js}
\end{figure}

To demonstrate that this TMX model can reproduce essential features of the FRTHs observed in experiments, we have shown in Fig. \ref{Js} different forms of $E(t)$ produced from the TMX model to mimic the effects of DIV. The figure shows that there are sub-bursts within these $E(t)$ and the number of sub-bursts decreases; similar late DIV FRTH shown in Fig. \ref{DIV}. In addition, the $ \tau_{_{IBI}} $ also decreases, agreeing with the experimental results. Note that the time scale of the $E(t)$ is comparable to that of the experiments. In order to generate Fig. \ref{Js}, we have assumed i) $J$ and $U$ increase with DIV as the cultures become mature and ii) the neurotransmitter re-cycling process becomes faster as the neuron niche improves \cite{taud}. Fig. \ref{Js}c are the time courses of $x$, $u$ and ${\chi_0}$. The figure shows that the interaction between $x$ and $u$ generates the sub-bursts while the depletion and the recovery of ${\chi_0}$ controls the stop and the start of the SB. Note that when $\tau_D$ is small (fast recycling), the sub-bursts can get very close and eventually disappear (Fig. \ref{Js}d). 

\begin{figure}[h]
	\begin{center}
		\includegraphics*[width=\columnwidth]{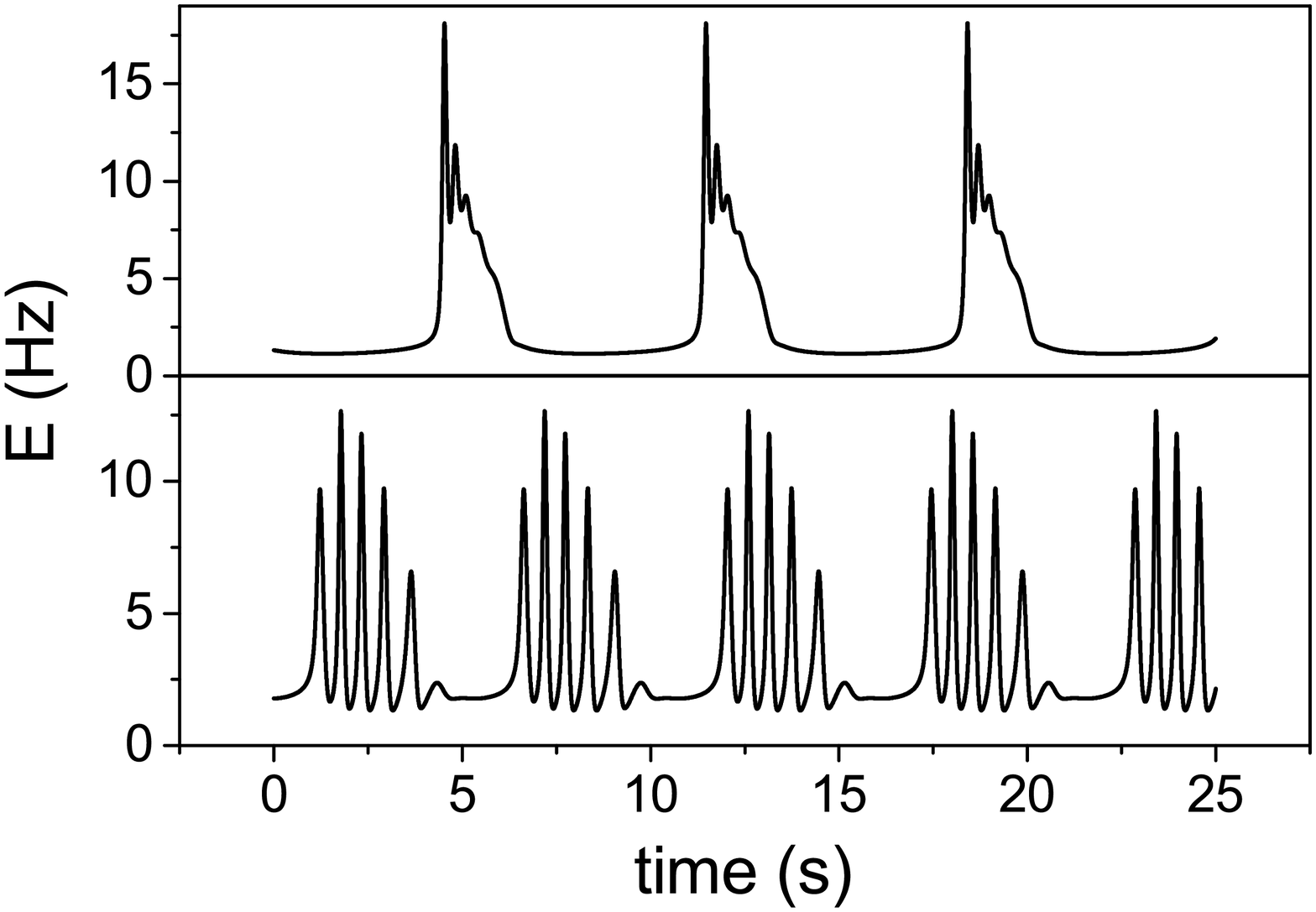}
	\end{center}
	\caption{ {\bf Time courses of $E(t)$ (in Hz) of the TMX model to mimic different magnesium treatment .} $E(t)$ (in Hz) from the TMX model for situations of a) normal  [Mg$ ^{2+} $] : $J=6.8$ and $\tau_D=0.1$ s and b) low  [Mg$ ^{2+} $] : $J=7.8$ and $\tau_D=0.15$ s. Other parameters are $\tau_X=20$ s, $I_0=-1.3$, $\beta=0.01$, $\tau=0.013$ s, $\tau_F=1.5$ s and $\alpha=1.5$.
		}
	\label{simuMg}
\end{figure}

To mimic the effects of  [Mg$ ^{2+} $]  on FRTH (Fig. \ref{treat}), we note that the unblocking effect of the NMDA receptors by a decrease in  [Mg$ ^{2+} $]  can be viewed as an increase in effective $J$ which will lead to an increase in firing rate. Since an increase in firing rate will make the recycling of the neurotransmitters slower \cite{kissandren}, the overall effects of a decrease in  [Mg$ ^{2+} $]  will be modeled in the TMX model as an increase in $J$ and an increase in $\tau_D$ as shown in the Fig. \ref{simuMg}. The inset shows that sub-bursts can indeed be induced by the lengthening of $\tau_D$ while an increase in $J$ shortens the $ \tau_{_{IBI}}$; similar to our experimental findings. 

\begin{figure}[h]
	\begin{center}
		\includegraphics*[width=\columnwidth]{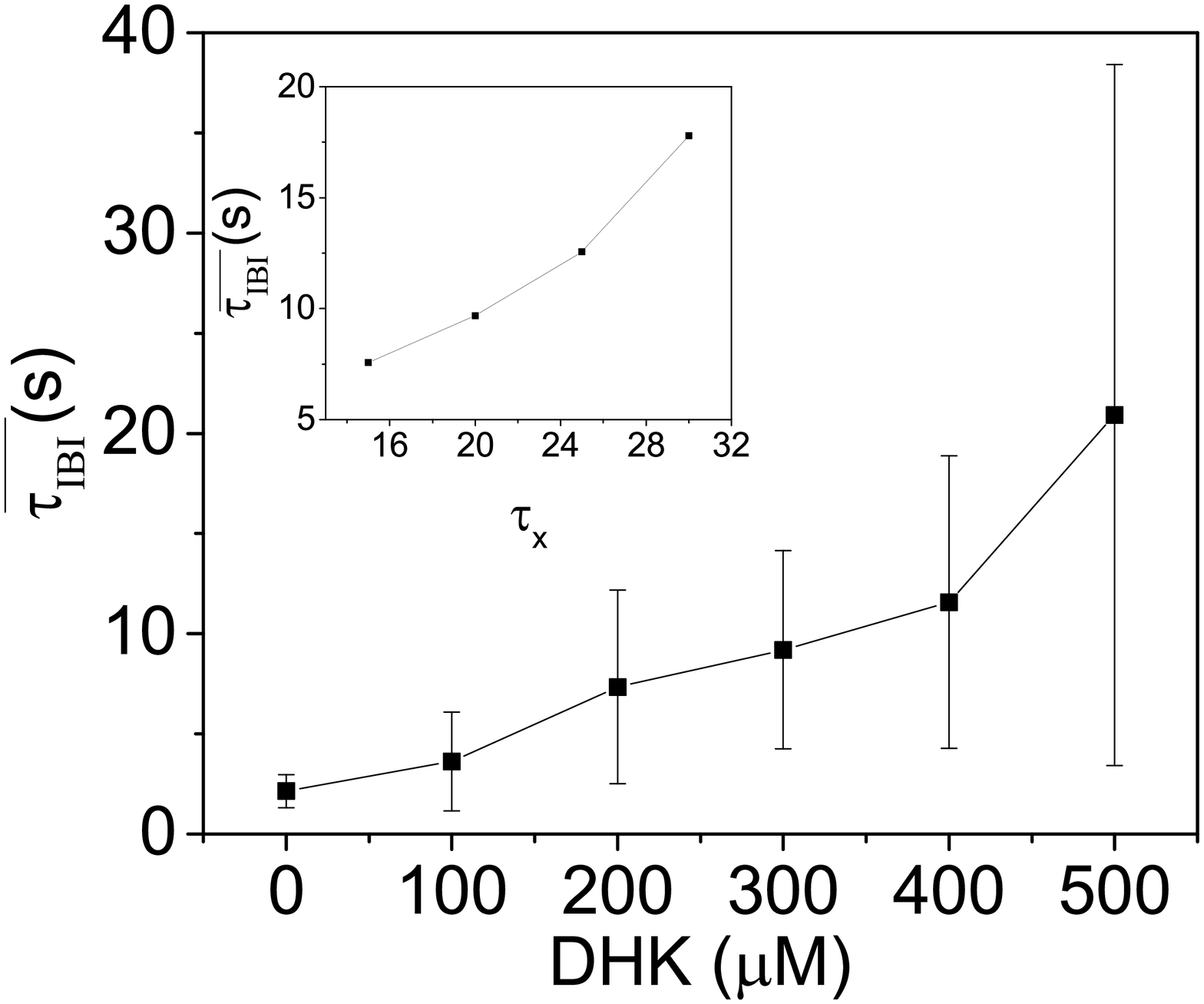}
	\end{center}
	\caption {{\bf Measurement of $\overline{\tau_{_{IBI}}}$ as a function of concentration of DHK (astrocyte glutamate uptake inhibitor) for a culture at 33 DIV}. Similar results can be observed 5 out of 6 samples from 3 dissections. The inset shows the dependence of $\overline{\tau_{_{IBI}}}$  on $ \tau_{X} $ in the TMX model. Other parameters are $X_0=0.95$, $ J =5.8$, $ U=0.3 $, $I_0=-1.3$, $\beta=0.01$, $\tau=0.013$ s, $\tau_D=0.15$ s, $\tau_F=1.5$ s and $\alpha=1.5$}. 
	\label{DHK}
\end{figure}

The success of the TMX model requires the existence of the modulations of available neurotransmitters (glutamate) $\chi_0$.  A known mechanism for this modulation is the uptake and recyling of glutamate from the astrocytes \cite{DHK1} surrounding the synapses. These astrocytes will convert the uptaken glutamate to glutamine and then transfer the glutamine back to a presynaptic neuron. The modulation of $\chi_0$ in the TMX model is most likely related to this process. To test this idea,  experiments are performed with dihydrokainate (DHK); an astrocytic glutamate transporter (GLT-1) blocker \cite{DHK1}. Since the blocking of GLT-1 will retard  the glutamate uptake, the blocking will produce an increase of $\tau_X $ in the TMX model. When $\tau_X$ is increased in TMX model, one will expected an increase in inter-burst interval  $\overline{\tau_{_{IBI}}}$.  In fact, this prediction is supported by our experiments with DHK as shown in Fig. \ref{DHK} which is a measurement of $\overline{\tau_{_{IBI}}}$ as a function of DHK concentration. The inset of Fig. \ref{DHK} is the corresponding simulation result from the TMX model. These results demonstrate that the TMX model is consistent with the recycling of glutamate by astrocytes.

\section*{Discussion}
Although the TMX model can capture some essential features of the dynamics of SBs observed in our experiments, the periodicity of $E(t)$ predicted are not seen in experiments. It is because the mean-field nature of the TMX model will work only when the system is synchronized. When the culture is in the low firing state before the occurrence of SB, the dynamics of the network cannot be described by a mean-field and the firing of a particular neuron in the system can triggered a SB. In this case, we do not expect to see the periodic generation of SBs.

Also, Eqn(4) of the TMX model to take into account the effects of glia is purely phenomenological. It is needed to stop the oscillations of the TM model by a depletion of available neuro-transmitters. The form of Eqn(4) is only one of many possible forms. Its dependence on $E$ and the empirical values of the constants $\tau_\chi$ and $\beta$ remained to be measured by further experiments. However, the main idea here is that we have given a concrete example to support the long standing hypothesis that glia can take part in the regulation of the overall activity of a neural network. In this case, the involvement of glia turns a non-stopping oscillation into bursts. 

\section*{Conclusion}
Several studies have shown that network connectivity and synaptic depression are critical parameters to control burst appearance \cite{simueq1}. Our TMX model is consistent with these findings if connections  and synapses efficacy increases in the network as the culture matures \cite{structureDIV}. However, the effect of neurotransmitter recycling in the TMX model, which has not been considered before, is essential for the understanding of experimental observations of sub-bursts within a burst. Furthermore, since the forms of bursts in the TMX model are controlled by the intrinsic parameters of the synapses, information written to the network will be wiped out by these endogenous bursts. In our view, these SB cannot be used to carry information of the system and perform computation as believed.  Since SBs can also be observed in an acute slice preparation from a functional brain only when the effective recurrent connections are artificially increased by the lowering of [Mg$ ^{2+} $] (a non-physiological condition \cite{lowmg}), the existence of SBs in a neuronal system might signal that it is perhaps in a pathological state such as epilepsy which is also characterized by synchronized firing over large area of the brain. 
The fact that the SBs are generated spontaneously in our 2D cultures under physiological  [Mg$^{2+} $] suggests that there might already be too many connections in the network and therefore not suitable for the study of normal functions of a neuronal system.

\section*{Acknowledgments}

This work has been supported by the MOST of ROC under the grant nos. NSC 100-2923-M-001-008-MY3, 101-2112-M-008-004-MY3 and NCTS of Taiwan. 

\nolinenumbers

%
%
%

\bibliographystyle{plos2015}

\bibliography{SBPLOS01}

\end{document}